The Future Of Mobile Security

Stefan Certic

https://www.certic.info/

**Introduction**

Mobile devices are more than just phones, they are a lifeline to the outdoor world, entertainment platform, GPS system, a little black book and a shopping and banking tool. What is not well known is that these devices are also gateways. Mobile devices can be used by a hacker as an access point into many other aspects of your digital life as well the lives of others in your network, making mobile security about more than just protecting your phone.

Since mobile phones are used key to your digital identity, hackers can use your mobile device as a way to get to other devices. This means that these hacks could not only lead to attacks on your other devices, but also the devices of anyone connected on the same network as you. A recent common hacking method works by forwarding the network traffic sent by the control server to another host in the network, which could be any other device inside a corporate network if the mobile device is connected to an internal Wi-Fi. Essentially, if you fall victim to this, you are exposing attackers to your entire digital life as well as that of anyone connected to your network. Worms and Man-in-the-Middle attacks are other examples of threats in which a hacker could potentially use one mobile device as the access point to other devices.

Mobile phones have traditionally been looked upon as a tool for making phone calls. But these days, given the developments in hardware and software, mobile phones use have been expanded to send messages, check emails, store contacts, store important dates, just to mention a few uses. Mobile connectivity options have also increased. After standard GSM connections, mobile phones now have 3G, 4G, LAN and WLAN connectivity. Most of us, if not all of us, carry mobile phones for communication purpose. Several mobile banking services are also now available to take gain from the improving capabilities of mobile devices. Today, mobile phones if configured are able to receive information on account balances in the form of SMS messages to using WAP and Java together with GPRS to allow fund transfers between accounts, stock trading, and confirmation of direct payments via the phone's micro browser. Installing both vendor-specific and third party applications allow mobile phones to provide these expanded new services other than communication.

However, today most security concerns on the rise are areas such as banks, governmental applications, health care industry, military organization, educational organizations, etc. Government organizations are setting standards, passing laws and forcing organizations and agencies to comply with these standards with non-compliance being met with wide-ranging consequences. There are several issues when it comes to security concerns in these numerous and varying industries with one common weak link being passwords.

Most systems today depend on static passwords to authenticate the user's identity. However, such passwords come with major management of security concerns that have been known to be exploited by hackers. Users have a tendency to to use easy-to-guess passwords, use the same password in numerous accounts, write the passwords or store them on their machines, etc. When users do this, hackers have the option of using many techniques to steal passwords such as shoulder surfing, snooping, sniffing, guessing, just to mention a few. Several 'proper' strategies for using passwords have been suggested. Some of the proposed strategies are very difficult to use and others might not meet the company's security concerns.

One of the ways that mobile devices security can be improved is through two-step authentication system. It consists of a server connected to a GSM enabled service provider and a mobile phone client equipped with SMS receiving functionality. This system involves an implementation where corporate server web application authorize customer with username / password, then connect with a service provider who will generate token and send token to customer via SMS returning transaction ID, asking customer to enter token, then match entered token against transaction id obtained from service provider.

Two step verification method by CS networks (M) Secure authentication provides an additional level of security to web application, registration system, or login procedure. On top of the username with password, user is required to validate a one-time only code that CS Networks two-step authentication will send via SMS text message. CS Network Two Way Authentication is designed for everyone who needs to protect themselves from unauthorized access by stealing password, social engineering and similar techniques of gaining access without proper permissions. It significantly decreases the chances of having the personal information of your customers taken by someone else. The key reason why it is difficult for hackers is because not only do they need to get your password and your username, but must also gain access of your mobile phone as well.

**Two-Step Authentication**

To better understand this system we start by defining authentication. It is the use of one or more mechanisms to prove that you are who you claim to be. Once the identity of the human or machine is validated, access is granted. There are three universally renowned authentication factors which exist today: what you know (e.g. passwords), what you have (e.g. ATM card or tokens), and what you are (e.g. biometrics). In recent times work has been done in trying other factors such as a fourth factor, which is based on the notion of vouching. Example of this is, somebody you know.

The two factor authentication is a mechanism which implements two of the above mentioned factors and is therefore considered stronger and more secure than the traditionally implemented one factor authentication system. Withdrawing money from an ATM machine utilizes two factor authentications; the user must possess the ATM card, i.e. what you have, and must know a unique personal identification number (PIN), i.e. what you know.

Passwords are well-known to be one of the easiest targets of hackers. Therefore, most organizations are searching for more secure strategies to protect their customers and employees. One of these secure methods is biometrics which are known to be very secure and are used in special organizations, but they are not used much in secure online transactions or ATM machines given the expensive hardware that is needed to identify the subject and the maintenance costs, etc. Instead, banks and companies are using tokens as a mean of two factor authentication.

A security token is a physical device that an authorized user of computer services is given to aid in authentication. It is also referred to as an authentication token or a cryptographic token. Tokens come in two formats: hardware and software. Hardware tokens are small devices which are small and can be conveniently carried. Some of these tokens store cryptographic keys or biometric data, while others display a PIN that changes with time. At any particular time when a user wishes to log-in, i.e. authenticate, he uses the PIN displayed on the token in addition to his normal account password. Software tokens are programs that run on computers and provide a PIN that change with time. Such programs implement a One Time Password (OTP) algorithm. OTP algorithms are critical to the security of systems employing them since un authorized users should not be able to guess the next password in the sequence. The sequence should be random to the maximum possible extent, unpredictable, and irreversible.

CS Networks Two-Step Authentication is a mobile-based software token system that once adopted would replace existing hardware and computer-based software tokens. The system is secure and consists of three parts: (1) client's mobile phone, (2) server software connected to a Service Provider, and (3) a Service Provider Itself.

In order for the server to verify the identity of the user, the user sends to the server, via a remote computer, information unique to the user. The server checks the message content and if correct, returns a randomly generated OTP to the mobile phone. The user will then have a given amount of time to use the OTP before it expires. Note that this method will require both the client and server to pay for the telecommunication charges of sending the SMS message.

**Components of a Two-Step Authentication:**

**A. OTP Algorithm**

In order to secure the system, the generated OTP must be hard to guess, retrieve, or trace by hackers. Therefore, it's very important to develop a secure OTP generating algorithm. Several factors can be used by the OTP algorithm to generate a difficult-to-guess password. Users seem to be willing to use simple factors such as their mobile number and a PIN for services such as authorizing mobile micropayments. Note that these factors must exist on both the mobile phone and server in order for both sides to generate the same password. In most two-step authentication design, the following factors are chosen:

• *IMEI number*: The term stands for International Mobile Equipment Identity which is unique to each mobile phone allowing each user to be identified by his device. This is accessible on the mobile phone and will be stored in the server's database for each client.

• *IMSI number*: The term stands for International Mobile Subscriber Identity which is a unique number associated with all GSM and Universal Mobile Telecommunications System (UMTS) network mobile phone users. It is stored in the Subscriber Identity Module (SIM) card in the mobile phone. This number will also be stored in the server's database for each client.

• *Username*: Although no longer required because the IMEI will uniquely identify the user anyway. This is used together with the PIN to protect the user in case the mobile phone is stolen.

• *PIN*: This is required to verify that no one other than the user is using the phone to generate the user's OTP. The PIN together with the username is data that only the user knows so even if the mobile phone is stolen the OTP cannot be generated correctly without knowing the user's PIN. Note that the username and the PIN are never stored in the mobile's memory. They are just used to generate the OTP and discarded immediately after that. In order for the PIN to be hard to guess or brute-forced by the hacker, usually, a minimum of 8-characters long PIN is requested with a mixture of upper- and lower-case characters, digits, and symbols.

• *Hour*: This allows the OTP generated each hour to be unique.

• *Minute*: This would make the OTP generated each minute to be unique; hence the OTP would be valid for one minute only and might be inconvenient to the user. An alternative solution is to only use the first digit of the minute which will make the password valid for ten minutes and will be more convenient for the users, since some users need more than a minute to read and enter the OTP. Note, that the software can modified to allow the administrators to select their preferred OTP validity interval.

• *Day*: Makes the OTP set unique to each day of the week.

• *Year/Month/Date*: Using the last two digits of the year and the date and month makes the OTP unique for that particular date.

The time is retrieved by the client and server from the telecommunication company. This will ensure the correct time synchronization between both sides.

For example, in most OTP algorithm the above factors are concatenated and the result is hashed using SHA-256 which returns a 256 bit message. The message is then XOR-ed with the PIN replicated to 256 characters. The result is then Base64 encoded which yields a 28 character message. The message is then shrunk to an administrator-specified length by breaking it into two halves and XOR-ing the two halves repeatedly. This process results in a password that is unique for a ten minute interval for a specific user. Keeping the password at 28 characters is more secure but more difficult to use by the client, since the user must enter all 28 characters to the online webpage or ATM machine. The shorter the OTP message the easier it is for the user, but also the easier it is to be hacked. The proposed system gives the administrator the advantage of selecting the password's length based on his preference and security needs.

### B. Client Design

For a two-step authentication, data transmission, in addition to the mobile identification number, are encrypted via a 256-bit symmetric key in the OTP algorithm and sent to the server. The server decrypts the message via the same 256-bit symmetric key, extracts the identification factors, compares the factors to the ones stored in the database, generates an OTP and sends the OTP to the client's mobile phone if the factors are valid. The advantage of encrypting the SMS message is to prohibit sniffing or man-in-the-middle attacks. The 256-bit key will be extremely hard to brute-force by the hacker. Note that each user will have a pre-defined unique 256-bit symmetric key that is stored on both the server and client at registration time.

### C. Database Design

A database is needed on the server side to store the client's identification information such as the first name, last name, username, password, unique symmetric key, and the mobile telephone number for each user. The password field will store the hash of the OTP password. It will not store the password itself. Should the database be compromised the hashes cannot be reversed in order to get the passwords used to generate those hashes. Hence, the OTP algorithm will not be traced.

### D. Server Design

In order for the CS Networks two-step authentication to function as expected a server is implemented to generate the OTP on the organization's side. The server consists of a database as described in Section above and is connected to a CS Networks Authentification Platform.

The server application is multi-threaded. The first thread is responsible for initializing the database and provider connection, listening for client requests. The second thread is responsible for verifying the SMS information, and generating and sending the OTP. A third thread is used to compare the OTP to the one retrieved using the connection-less method.

In order to setup the database, the client must register in person at the organization. The client's mobile phone and username are retrieved and stored in the database, in addition to the passwor. Once the server is prompted to undertake an operation and the details (username, password and phone identification number) are correct, it automatically sends an OTP to the phone via CS Networks Two-Factor platform after which verification takes place.

### Summary of how CS Networks Two Step Authentication works

This a summary of how CS Networks Two Step Authentication works explained in steps:

I. Your application sends request to our server with a mobile number as a parameter.

II. CS Networks application returns Unique Transaction ID and sends randomly generated Token to a given number.

III. Customer needs to enter received token.

IV. Your application validate entered token against Transaction ID, and get response from CS Networks server whether the token is Valid or not. Ether approving or denying access to sensitive data.

Unlike other Two Step Authentication, our system is easier to use and is economical in terms of cost. Data is spread via multiple systems, making it impossible to gain access without exploiting both Corporate Network, CS Networks as a provider, destination device itself.

### Future of CS networks two-step authentication

There has been a lot of debate lately surrounding Two Factor Authentication. I have been watching and participating in this debate between two camps. One camp of enthusiasts who have confidence in the Two Factor Authentication as the future of security measures and a second camp which seem to be certain that because the technology has been proven to be susceptible, that it will slowly die away and give way to a newer more all-inclusive security technology.

Am with the camp of those who have faith in two step authentication, and specifically the CS networks implementation mechanism, to be the best type of security we currently have. The technology, when used correctly online can help make major sites such as banking and credit sites much more secure than they previously were when only a password was necessary. Major companies like Microsoft have even begun implementing various types of Two Factor Authentication into various parts of their business. Further demonstrating that the technology works, PayPal made use of this technology recently and has incorporated it into their web services. This has added a layer of security to their services that can't be found with many other similar services thus giving them an edge in a very competitive market.

Even with these companies adopting Two Factor Authentication and though the technology continues to improve there are many, pessimist all of whom claim that the technology is not effective and that it's not perfect and has been compromised in the past. To some extent their concern is valid but as I have outlined the benefits above, the benefits outweigh the cost. All programs have been circumvented at some point, but with additional security layers you reduce this possibility. Passwords are an outdated and highly susceptible technology, however, many websites, and programs may not require more security due to low level of sensitivity of data stored in them.

For CS Networks Two Step Authentication, I have always felt that while it may be an imperfect technology so are ninety nine percent of all others. With that in mind I also feel that it may be misunderstood. Certainly many vendors and those in favor of the technology tout it as the be all and end all to security technology which makes it an easy target for detractors looking to tone down its prominence and benefits. Our Two Step Authentication is easier to use and its also not as expensive as other Two Step Authentication systems.

In closing I'd like to add that while it has its issues, I see CS Network Two Factor Authentication becoming an industry standard in a short amount of time. Things like biometrics and Authentication 2.0 further my feelings on the future of this security. With so many businesses adopting TFA, it's only a matter of time before other businesses follow suit.

### Benefits of CS Networks Two-Step Authentication

### I. Reduced Risk

Single factor authenticating uses only one form of ID which is prone to hacking. Due to the technology available today, almost any single form of authentication can be fabricated. In many cases, such as with brute force attacks where passwords are guessed, cases of hack are ignored because a valid user might justifiably forget his password and might make several attempts to enter it correctly. If a company introduces a second level of authentication through CS networks (M) Secure, the level of complexity increases considerably. The higher the complexity of authorization, the higher the probability that the hacker will be caught. The greater chance of getting caught, the less likely that an attempt will be made by a hacker to break in to a computer or network. This results into a reduced risk of loss for the company.

### II. Minimize time consumed by increasing efficiency

Passwords might seem like a simple notion, but when classified and confidential resources are at risk, there is much more to a password than making a thief guess your birth day or child's name. Some four-digit passwords are far more popular than others: "1234" alone accounts for almost 11 percent of these passwords;

"1111," an additional 6 percent. Repetitive patterns occupy many of the other spots among the 20 most frequent numbers. Lower on the list are numbers that are likely to be a year of birth or the four-digit rendering of the month and day of a birthday. There must be complexity. The password complexity requirements outlined in your company's security policy might be common knowledge for your network administrators who manage the network, but few users are typically aware of these policies and requirements. This leads to weak passwords, which can be easily guessed or cracked, and security risks.

To prevent this, administrators will send out emails to users and post notices in an attempt to educate users. The sheer number of help desk calls of users who have forgotten their password is a testament to the futility of these endeavors. Having a strong, two factor authentication allows for a simple, automated, and intuitive process that users can get behind without getting bent out of shape from complex passwords. Instead of passwords, swipe cards can be used for the first authentication factor, and a thumb print for the second authentication factor. The help desk is then freed to do more than reset passwords and the network administrators can get back to working on the network.

### III. Enhanced Security

A password can be casually written on a piece of paper and left in plain sight for all to see. That password can then be used by an unscrupulous individual to gain access to restricted resources. Is the person who is using the password the intended user? It's impossible to tell. However, a CS Networks (M) Secure system second authentication factor enhances security by introducing an independent type of ID, one only the original person should be able to provide. No longer can a hacker hide behind an anonymous password. They must also provide physical proof to verify identity or they are denied access.

### IV. Resistant to Password Compromise

CS Networks Two factor authentication mitigates the risks inherent in the use of passwords. Passwords have a number of problems associated with them. If you choose a secure password, it can be difficult to remember--especially if you use a different password for every account you use. You may be tempted to write it down or give it to someone else for safe keeping. It may also be tempting to select an insecure password. These are easy to remember, but also frequently easy to guess or crack. Alternately, you may choose a secure password but end up using it everywhere. Once a single account is compromised, all of your accounts become insecure. By adding a second layer of security, two factor authentication helps keep intruders out of your accounts. Even if someone has your password, they are unable to do any damage without your key--whether this is a fingerprint, randomly generated number which changes each minute or private encryption key.

### V. Digital Trust

When communicating with others online, it may be difficult to prove your identity. This is especially true if you are communicating pseudonymously or with people you do not know very well. CS Networks Two factor authentication establishes a sense of digital trust.

### Conclusion

CS Network Two Way Authentication is designed for everyone who needs to protect themselves from unauthorized access by stealing password, social engineering and similar techniques of gaining access without proper permissions. 2-step authentication significantly decreases the chances of having the personal information of your customers taken by someone else. The key reason why is simply because hackers must not only get their password and your username, but a control of their mobile phone as well. CS Networks (M)Secure suite can be configured to perform additional checks to determinate possible fraud score based mobile physical location and state, as well to detect forwarded / possibly spoofed numbers. Advanced utilities of (M)Secure are developed for banking and financial insurance institutions to provide highest possible level of security and fraud prevention. I see CS Networks Two Step Authentication as the future of two factor authentication due to the fact that its simple to use and also economical to install.